# Exact Solution for Oscillations of Two Neutrinos within Alternative Theory of Neutrino Oscillations

Viliam Pažma[1], Julius Vanko[2] and Juraj Chovan[2]

[1]Institute of Physics, Comenius University, Bratislava, Slovakia
[2]Department of Nuclear Physics, Comenius university, Bratislava, Slovakia

vanko@fmph.uniba.sk

We present one point of contact between the standard theory of neutrino oscillations and the alternative one in which flavor neutrinos are described by states with definite masses. We show that both theories give the same results for static neutrinos only.

1. Introduction

The title of this paper means that we shall consider the existence of two kinds of neutrinos only and let it be for instance $\nu_1 = \nu_e$ and $\nu_2 = \nu_\mu$ and we shall study the transitions (oscillations) $\nu_1 \leftrightarrow \nu_2$. These oscillations will be studied in the framework of the following theory [1]. The free neutrinos $\nu_1, \nu_2$ are described by the hamiltonian

$$H_0 = \vec{\alpha}.\vec{p} + \beta M_d , \qquad (1)$$

where $M_d = diag\,(m_1, m_2)$, $m_1\,(m_2)$ is the mass of $\nu_1\,(\nu_2)$ and the standard meaning of other symbols is assumed.

The transitions $\nu_1 \leftrightarrow \nu_2$ will occur in the theory if the time development of states of $\nu_1, \nu_2$ will be determined by the hamiltonian

$$H = \vec{\alpha}\vec{p} + \beta \begin{pmatrix} m_1 & q \\ q^* & m_2 \end{pmatrix} . \qquad (2)$$

For a simplicity we shall assume $q^* = q$. (The case $q^* \neq q$ does not represent any serious problem.)

The eigenvalues of

$$M = \begin{pmatrix} m_1 & q \\ q & m_2 \end{pmatrix} \qquad (2')$$

are

$$M_1 = \tfrac{1}{2}(m_2 + m_1) - \tfrac{1}{2}\sqrt{(m_2 - m_1)^2 + 4q^2},$$

$$M_2 = \tfrac{1}{2}(m_2 + m_1) + \tfrac{1}{2}\sqrt{(m_2 - m_1)^2 + 4q^2}. \qquad (3)$$

It is evident that $M_1$ may be negative and thus (2) has not to describe two free Dirac particles.
In the next we shall calculate the probability $P(\nu_1 \leftrightarrow \nu_2; t)$ of the transition $\nu_1 \leftrightarrow \nu_2$ in time $t$. We shall work within the standard representation of Dirac matrices and so we choose

$$\vec{\alpha} = \begin{pmatrix} 0 & \vec{\sigma} \\ \vec{\sigma} & 0 \end{pmatrix}, \qquad \beta = \begin{pmatrix} 1 & 0 \\ 0 & -1 \end{pmatrix}.$$

2. Eigenvalues an eigenfunctions of $H_0$ and $H$

Let us now consider the equation

$$\varepsilon \varphi = H_0 \varphi \qquad (4)$$

with $H_0$ given by (1). If we are interested in neutrinos $\nu_1, \nu_2$ having the momentum $\vec{p}$ then we have to take into account only those solutions to (4) for which $\varepsilon > 0$ and

$$\vec{\Sigma} \cdot \vec{p}\varphi = -|\vec{p}|\varphi, \qquad \left(\vec{\Sigma} = \begin{pmatrix} \vec{\sigma} & 0 \\ 0 & \vec{\sigma} \end{pmatrix}\right). \qquad (5)$$

If we chose $\vec{p} = (0, 0, p > 0)$ then the mentioned above solutions are

$$\varphi_1 = \begin{pmatrix} A_1 \\ 0 \end{pmatrix} \quad \text{for } \varepsilon_1 = \sqrt{p^2 + m_1^2},$$

$$\varphi_2 = \begin{pmatrix} 0 \\ A_2 \end{pmatrix} \quad \text{for} \quad \varepsilon_2 = \sqrt{p^2 + m_2^2}, \tag{6}$$

where

$$A_i = \begin{pmatrix} \sqrt{\varepsilon_i + m_i}\, w \\ -\sqrt{\varepsilon_i - m_i}\, w \end{pmatrix} \quad \text{and} \quad w = \begin{pmatrix} 0 \\ 1 \end{pmatrix}. \tag{7}$$

The solutions $\varphi'_s$ corresponding to $\varepsilon < 0$, momentum $-\vec{p}$ and the positive helicity we choose in the form

$$\varphi_3 = \begin{pmatrix} A_3 \\ 0 \end{pmatrix} \quad \text{for} \quad \varepsilon_3 = -\sqrt{p^2 + m_1^2} = -\varepsilon_1,$$

$$\varphi_4 = \begin{pmatrix} 0 \\ A_4 \end{pmatrix} \quad \text{for} \quad \varepsilon_4 = -\sqrt{p^2 + m_2^2} = -\varepsilon_2, \tag{8}$$

where

$$A_i = \begin{pmatrix} -\sqrt{|\varepsilon_i| - |m_i|}\, w \\ \sqrt{|\varepsilon_i| + |m_i|}\, w \end{pmatrix}, \quad (i = 3, 4) \tag{9}$$

and $m_3 = -m_1$, $m_4 = -m_2$.

The $\varphi_i, (i = 1, 2, 3, 4)$ are governed by

$$\overline{\varphi}_i \varphi_i = 2 m_i \tag{10}$$

and satisfy

$$\sum_{i=1}^{4} \frac{\varphi_i \overline{\varphi}_i}{2 m_i} = diag\, (01010101) \tag{11}$$

Because we deal with solutions $\varphi$'s of the type $\varphi^+ = (0\ a_1 0\ a_2 0\ a_3 0\ a_4)$ and their linear combinations then $I'_8 = diag\,(01010101)$ is the unit operator acting on the considered subspace, i.e., $\varphi^+ I'_8 = \varphi^+$, $I'_8 \varphi = \varphi$.

Let us now look for eigenvalues and eigenfunctions of $H$. We start with the equation

$$E \Phi = H \Phi \qquad (12)$$

and consider solutions with negative (positive) helicity for $E > 0, \vec{p}$ ($E < 0, -\vec{p}$).
Putting

$$\Phi = \begin{pmatrix} A' \\ B' \end{pmatrix}$$

then we get from (12)

$$(\vec{\alpha}\vec{p} + \beta m_1)A' + \beta q B' = EA',$$

$$(\vec{\alpha}\vec{p} + \beta m_2)B' + \beta q A' = EB'. \qquad (13)$$

The eigenvalues of $H$ are

$$E_i = \pm\sqrt{\vec{p}^2 + M_i^2}, \qquad (i = 1, 2)$$

where $M_i$ are given by (3).
Confining ourselves to $\vec{p} = (0, 0, p > 0)$, $E_i = \sqrt{p^2 + M_i^2}$ and negative helicity than

$$\Phi_i = \frac{1}{\sqrt{1 + \xi_i^2}} \begin{pmatrix} A_i' \\ \xi_i A_i' \end{pmatrix}, \qquad (i = 1, 2) \qquad (14)$$

where

$$A_i' = \begin{pmatrix} \sqrt{E_i + M_i}\, w \\ -\sqrt{E_i - M_i}\, w \end{pmatrix}, \qquad \xi_i = \frac{M_i - m_i}{q}. \qquad (15)$$

The solutions $\Phi_i (i = 3, 4)$ to (12) corresponding to $E_i = -\sqrt{p^2 + M_i^2}, -\vec{p}$ and the positive helicity are

$$\Phi_i = \frac{1}{\sqrt{1+\xi_i^2}} \begin{pmatrix} A_i' \\ \xi_i A_i' \end{pmatrix}, \quad (i=3,4) \tag{16}$$

where

$$A_i' = \begin{pmatrix} -\sqrt{|E_i|-|M_i|}\, w \\ \sqrt{|E_i|+|M_i|}\, w \end{pmatrix}, \quad \xi_3 = \xi_1, \xi_4 = \xi_2, \tag{17}$$

and $M_3 = -M_1$, $M_4 = -M_2$.

The solutions $\Phi_i (i=1,2,3,4)$ satisfy

$$\overline{\Phi_i}\,\Phi_i = 2M_i \qquad \sum_{i=1}^{4} \frac{\Phi_i \overline{\Phi_i}}{2M_i} = diag\,(01010101) \tag{18}$$

3. Amplitude of transition $\nu_1 \leftrightarrow \nu_2$ in time $t$

Having the initial state $\varphi_1/\sqrt{2\varepsilon_1}$ in time $t=0$ then its time-development is given by the equation

$$\Phi = e^{-itH} \frac{1}{\sqrt{2\varepsilon_1}} \varphi_1. \tag{19}$$

On the basis of (11) and (18) we can write

$$\varphi_i = \sum_{j=1}^{4} \frac{\Phi_j \overline{\Phi_j}}{2M_j} \varphi_i = \sum_{j=1}^{4} W_{ij} \Phi_j$$

where

$$W_{ij} = \frac{1}{2M_j} \overline{\Phi_j} \varphi_i$$

and

$$\Phi_i = V_{ij} \varphi_j$$

where

$$V_{ij} = \frac{\overline{\varphi}_j \Phi_i}{2m_j} = (W^{-1})_{ij}.$$

Now (19) can be rewritten into the form

$$\Phi = W_{1j}\, e^{-iE_j t}\, V_{jk}\, \sqrt{\frac{|\varepsilon_k|}{\varepsilon_1}}\, \frac{\varphi_k}{\sqrt{2|\varepsilon_k|}} \qquad (20)$$

and the amplitude $A(v_1 \to v_k; t)$ of the transition $v_1 \to v_k$ in time $t$ is equal to

$$A(v_1 \to v_k; t) = W_{1j}\, e^{-iE_j t}\, V_{jk}\, \sqrt{\frac{|\varepsilon_k|}{\varepsilon_1}}. \qquad (21)$$

We note that $A(v_1 \to v_k; t)$ for $k = 1, 2$ corresponds to transitions $v_1 \to v_1$, $v_1 \to v_2$ and for $k = 3, 4$ corresponds to $v_1 \to \tilde{v}_1$ (antineutrino), $v_1 \to \tilde{v}_2$.

4. The relation between the standard theory and the considered one

Let us now look for some point of contact between (21) and the following expression for A which follows from the standard theory
( see e.g. [2 - 6] )

$$A(v_1 \to v_k; t) = U_{1j}\, e^{-iE_j t}\, U^{+}_{jk}, \qquad (k = 1, 2), \qquad (22)$$

where $U$ is the unitary matrix defining states $|v_i\rangle$ of the flavor neutrinos $v_i$ by means of states $|v'_i\rangle$ of mass neutrinos $v'_i$ by the equation $|v_i\rangle = U_{ij}|v'_j\rangle$. The states $|v'_i\rangle$ for $\vec{p} = 0$ are eigenstates of the mass matrix $M$ and thus $UMU^{+} =$ diagonal matrix.
Putting

$$W = \begin{pmatrix} W^{(1)} & W^{(2)} \\ W^{(3)} & W^{(4)} \end{pmatrix}, \qquad V = \begin{pmatrix} V^{(1)} & V^{(2)} \\ V^{(3)} & V^{(4)} \end{pmatrix}$$

where $W^{(\alpha)}, V^{(\alpha)} (\alpha = 1, 2, 3, 4)$ are $2 \times 2$ matrices then

$$W_{ij}^{(1)} = \frac{\xi_j^{i-1}}{2M_j\sqrt{1+\xi_j^2}} \left[ \sqrt{(\varepsilon_i + m_i)(E_j + M_j)} - \sqrt{(\varepsilon_i - m_i)(E_j - M_j)} \right] = W_{ij}^{(4)},$$

$$W_{ij}^{(2)} = \frac{\xi_j^{i-1}}{2M_j\sqrt{1+\xi_j^2}} \left[ \sqrt{(\varepsilon_i + m_i)(E_j - M_j)} - \sqrt{(\varepsilon_i - m_i)(E_j + M_j)} \right] = W_{ij}^{(3)},$$

$$V_{ij}^{(1)} = \frac{\xi_i^{j-1}}{2m_j\sqrt{1+\xi_i^2}} \left[ \sqrt{(E_i + M_i)(\varepsilon_j + m_j)} - \sqrt{(E_i - M_i)(\varepsilon_j - m_j)} \right] = V_{ij}^{(4)},$$

$$V_{ij}^{(2)} = \frac{\xi_i^{j-1}}{2m_j\sqrt{1+\xi_i^2}} \left[ \sqrt{(E_i + M_i)(\varepsilon_j - m_j)} - \sqrt{(E_i - M_i)(\varepsilon_j + m_j)} \right] = V_{ij}^{(3)}.$$

For $\vec{p} = 0$ we have

$$W_{ij}^{(1)} = \frac{\xi_j^{i-1}}{\sqrt{1+\xi_j^2}} \sqrt{\frac{m_i}{M_j}} = W_{ij}^{(4)},$$

$$W_{ij}^{(2)} = W_{ij}^{(3)} = 0,$$

$$V_{ij}^{(1)} = \frac{\xi_i^{j-1}}{\sqrt{1+\xi_i^2}} \sqrt{\frac{M_i}{m_j}} = V_{ij}^{(4)},$$

$$V_{ij}^{(2)} = V_{ij}^{(3)} = 0.$$

Now from (21) we get

$$A(\nu_1 \to \nu_k; t)|_{\vec{p}=0} = U_{1j} e^{-iM_j t} U_{jk}^{-1}, \qquad (k = 1, 2), \tag{23}$$

$$A(\nu_1 \to \nu_k; t)|_{\vec{p}=0} = 0 \qquad \text{for} \qquad (k = 3, 4),$$

where

$$U = \begin{pmatrix} \dfrac{1}{\sqrt{1+\xi_1^2}} & \dfrac{1}{\sqrt{1+\xi_2^2}} \\ \dfrac{\xi_1}{\sqrt{1+\xi_1^2}} & \dfrac{\xi_2}{\sqrt{1+\xi_2^2}} \end{pmatrix}$$

Now it is not difficult to show that (23) for $k = 1, 2$ and (22) are the same results. Namely, taking into account

$$M_1 M_2 = m_1 m_2 - q^2, \qquad M_1 + M_2 = m_1 + m_2$$

then it is not difficult to prove $\xi_1 \xi_2 = -1$ and then verify the following relations

$$U^+ U = U U^+ = I_2, \quad (I_2 \text{ is the unit } 2 \times 2\text{-matrix}),$$

$$U M U^+ = diag\,(M_1 M_2).$$

Hence, (23) indeed reduces at $\vec{p} = 0$ to (22).

Thus one can conclude that for static neutrinos $(\vec{p} = 0)$ the considered theory and the standard one give the same result.

## 5. Concluding remarks

Throughout this paper we worked with $M_1, M_2 > 0$ although one can admit $M_1 < 0$. Naturally, the previous calculations can be repeated without any difficulties also for $M_1 < 0$, $M_2 > 0$. We did not do so because we looked for some points of contact between the theories in question. Namely, the states $\Phi_i$'s play the same role in the presented theory as the mass neutrino states in the standard theory and the masses of the mass neutrinos are intuitively assumed to be positive.

We conjecture that the point of contact we found is the only one. The physical distinctions of these theories are evident (e.g. in this theory any neutrino $\nu_k$ ($k = 1, 2, 3$) is described by the state with definite mass). The contemporary experimental data are not sufficiently strong for rejecting one of these theories. Thus we are reffered on the next investigations which perhaps will reveal weak points of our speculations.

This work was supported by the Slovak grant agenture VEGA, Contract No.1/8315/01.

References

[1] V.Pažma, J.Vanko, hep-ph/0311090.
[2] S.M.Bilenky, hep-ph/0210128, hep-ph/037186.
[3] W.M.Alberico, S.M.Bilenky, hep-ph/0306239.
[4] A.Bellerive, hep-ex/0312045.
[5] W.Grimus, hep-ph/0307149.
[6] C.Giunti, M.Laveder, hep-ph0301276, hep-ph/0310238.